\documentclass[draft,ms]{article}

\usepackage[dvips]{graphicx}
 \usepackage[]{natbib}
\setkeys{Gin}{draft=false}
\begin{document}

%
%

\title{Estimating the compressibility of osmium from recent measurements of Ir-Os alloys under high pressure}
%
%

%
%



\author {N. V. Sarlis and
E. S. Skordas, \\
Department of Solid State Physics and\\
 Solid Earth Physics
Institute\\
 Faculty of Physics, School of Science, \\
 National and Kapodistrian University of Athens,\\
Panepistimiopolis, Zografos 157 84,\\
 Athens, Greece}
\maketitle

\begin{abstract}
Several fcc- and hcp-structured Ir-Os alloys have been recently
studied up to 30 GPa at room temperature by means of
synchrotron-based X-ray powder diffraction in diamond anvil cells.
Using their bulk moduli -which increase with increasing osmium
content showing a deviation from linearity- and after employing a
thermodynamical model it was concluded that the bulk modulus for
osmium is slightly smaller than that for diamond. Here, a similar
conclusion is obtained upon employing an alternative model, thus
strengthening the conclusion that osmium is the densest but not
the most incompressible element. This is particularly interesting for Earth Sciences
since it may be of key importance towards clarifying the anomalous elastic
properties of the Earth's core.
\end{abstract}

%
%

%


%
%

\section{Introduction}
\label{intro}
Heavy transition metals such as Ta, W, Re, Os and Ir have
exceptional mechanical, thermal and chemical stabilities. Their
very low compressibilities $\kappa$ make them essential subjects
for high pressure studies. The compressibility of materials, in
general, is of major importance because of its correlation to
material strength, chemical bonding, and electronic structure.
Strongly bonded materials have short interatomic distances and
correspondingly strong repulsive forces, resulting in high bulk
moduli $B (=1/\kappa)$. In this regard, the finding by
 \citet{PhysRevLett.88.135701} that the $B$ value for Os at ambient
conditions is $B=462(12)$GPa, i.e., larger than that for diamond
($B=443$GPa, see \citet{BOs}),  attracted major interest. Such a  large
$B$ value has implications for the nature of the metallic bond in
Os, where the bonding electrons are delocalized compared to the
covalently bonded diamond in which they are localized. Later
studies, however, came out with markedly lower $B$ values for Os.
In particular, values around 400GPa were reported by
 \citet{BOs}, \citet{TAK04} and \citet{GOD12}, while \citet{VOR05} found $B=435(19)$GPa
and \citet{ARM10} $B=421(3)$GPa. Hence, controversial values
concerning the bulk modulus for Os have been published to date.
This is challenging in view of the following: Osmium is the
densest element at ambient conditions, thus if its corresponding
$B$ value exceeds that of diamond (as reported by
 \citet{PhysRevLett.88.135701}), this would reflect that Os is also
the most incompressible element. In addition, osmium is of
particular importance in Earth Sciences because it provides key
information about the properties of iron and its alloys making up
Earth's core: The stability of hcp osmium is similar to that of
hcp iron, and a systematic characterization of these analog
materials will help towards understanding the anomalous properties
of Earth's core \citep{MORE86,CRE92}. The melting point of osmium
has been reported to be 3306K \citep{REI64,OKA94,VAC54,GEB63}.

An interesting study of the compressibility of Ir-Os alloys under
high pressure just appeared by \citet{Yusenko2015155}. As they
noticed phase stability in the Ir-Os system has importance for the
genesis of the natural Ir-Os-Ru and other pure platinum group
based metallic minerals which show complex nature and
non-equilibrium character at ambient conditions due to their
formation under high-pressure from poly-component melt in Earth's
mantle \citep{BIR80,FON12,WEI08}. They prepared several fcc- and
hcp- structured Ir-Os alloys and the measurement of the
corresponding atomic volumes at ambient conditions using powder
X-ray diffraction showed an almost linear dependence as a function
of composition. These alloys were studied up to $P=30$GPa at room
temperature by means of synchrotron-based X-ray powder diffraction
in diamond anvil cells and their bulk moduli were found to
increase with increasing osmium content showing a deviation from
linearity. This concentration dependence of bulk moduli was
shown by \citet{Yusenko2015155} to be satisfactorily described by a
thermodynamic model developed by \citet{VAR80AA}. According
to this model the bulk moduli $B_0 (x_{{\rm Os}})$ as a function
of the atomic fraction of osmium, $x_{{\rm Os}}$, is given by
\begin{equation} \label{eq1}
B_0 (x_{{\rm Os}})=B_{{\rm Ir}} \left[ \frac{1+x_{{\rm
Os}}\left(\frac{V_{{\rm Os}}}{V_{{\rm Ir}}}-1\right)}{1+x_{{\rm
Os}}\left(\frac{B_{{\rm Ir}}V_{{\rm Os}}}{B_{{\rm Os}}V_{{\rm
Ir}}}-1\right)} \right],
\end{equation}
where $B_{{\rm Ir}}$ and $B_{{\rm Os}}$ are bulk moduli and
$V_{{\rm Ir}}$ and $V_{{\rm Os}}$ are atomic volumes at ambient
pressure of pure Ir and Os, respectively. \citet{Yusenko2015155},
upon using their measured bulk moduli of the Ir-Os alloys and
employing Eq.(\ref{eq1}),  found the values 354(2)GPa and
442(4)GPa for pure Ir and Os, respectively. In particular,
\citet{Yusenko2015155} concluded that the bulk modulus for pure Os
is slightly smaller than that of the diamond. It is the scope of
the present short paper to investigate the validity of this
conclusion of \citet{Yusenko2015155} by using an alternative model
which interrelates the bulk modulus of an alloy with the bulk
moduli of its pure constituents. The basic idea of this model is
that when replacing an atom of a host crystal by a ``foreign''
atom, the corresponding volume variation can be considered as a
``defect volume''.

\section{The alternative model for the compressibility of a solid solution}\label{Sec2}
Let us explain how the compressibility $\kappa$ of a solid
solution A$_x$B$_{1-x}$ is interrelated with compressibilities of
the two end members A and B by following
\citet{VARBOOK} and \citet{SKO12}. We call the two end members A and B as
pure components 1 and 2, respectively, and label $v_1$ the volume
per atom of the pure component 1 and  $v_2$ the volume per atom of
the pure component 2. Let $V_1$  and $V_2$ denote the
corresponding molar volumes, i.e. $V_1=Nv_1$  and $V_2=Nv_2$
(where $N$ stands for Avogadro's number) and assume that $v_1 <
v_2$. We now define a ``defect  volume'' \citep{VARBOOK,SKO12} as
the increase of the volume $V_1$, if one atom of type 1 is
replaced by one atom of type 2. It is evident that the addition of
one ``atom'' of type 2 to a crystal containing  atoms of type 1
will increase its volume by $v_d+v_1$ (see pp.325 and 326 of
\citet{VARBOOK} as well as \citet{SKO12}). Assuming
that $v^d$ is independent of composition, the volume $V_{N+n}$ of
a crystal containing $N$ atoms of type 1 and  $n$ atoms of type 2
can be written as
\begin{equation}\label{eq2}
V_{N+n}=Nv_1+n(v^d+v_1) \Longleftrightarrow V_{N+n}= \left[
1+\frac{n}{N} \right] V_1+nv^d .
\end{equation}
The molar fraction $x$ is connected to $n/N$ by (see Eq.(12.5) on
page 328 of \citet{VARBOOK})
\begin{equation}\label{eq3}
\frac{n}{N}=\frac{x}{1-x} .
\end{equation}
The compressibility $\kappa$ of the solid solution (as well as its
bulk modulus $B=1/\kappa$) can be found by differentiating
Eq.(\ref{eq2}) with respect to pressure, which finally gives:
\begin{equation}\label{eq4}
\kappa V_{N+n}=\kappa_1V_1+\frac{n}{N} \left[ \kappa^d N v^d
+\kappa_1 V_1 \right],
\end{equation}
where $\kappa^d$ denotes the compressibility of the volume $v^d$,
defined as
\begin{equation}\label{eq5}
\kappa^d\equiv \frac{1}{B^d}=-\frac{1}{v^d} \left( \frac{\partial
v^d}{\partial P} \right)_T.
\end{equation}
The ``defect volume'' $v^d$ can be approximated by (see p.342 of
\cite{VARBOOK})
\begin{equation}\label{eq6}
v^d=\frac{V_2-V_1}{N}=v_2-v_1.
\end{equation}
Obviously $V_{N+n}$ can be obtained versus the composition by
means of Eq.(\ref{eq2}) when considering also Eq.(\ref{eq6}).
Then, the compressibility $\kappa$ can be studied versus the
composition from Eq.(\ref{eq4}) by assuming -to a first
approximation- that the compressibility $\kappa^d$ is independent
of composition. A rough estimation of $\kappa^d$ can be made by
employing a thermodynamical model \citep{VAR77A,VAR78B} for the
formation and migration of the defects in solids (cf. the
replacement of a host atom with a ``foreign'' one can be
considered in general as a defect \citep{PhysRevB.9.1866,VAR74,VAR74B}). This
model has been successfully applied to various categories of
solids including diamond \citep{PhysRevB.75.172107}, oxides \citep{Chroneos20151}, semiconductors \citep{Chroneos2015179}, silicates \citep{GGGE:GGGE20681} 
metals \citep{PhysRevB.24.3606,PSSB:PSSB2221020157}, ionic
crystals \citep{doi:10.1080/01418618008239352},
fluorides \citep{VAR819,VARALEX80,PSSA:PSSA2210880250,Varotsos2008438},
mixed alkali halides \citep{PSSB:PSSB2221000252,VAR81,VAR80B,VARO78} as well as
complex ionic materials under uniaxial stress that emit electric
signals before fracture \citep{VAR99,VAR92}, thus explaining the
signals detected before major
earthquakes \citep{VAR06B,NAT09V,NAT08,CHAOS2010,NEWTSA}. Within the
frame of this thermodynamical model, which states that the defect
Gibbs energy $g$ is proportional to the bulk modulus as well as to
the mean volume per atom, we first obtain the defect volume
$v[=(\partial g /\partial P)_T]$, and therefrom the
compressibility $\kappa^d$ (see Eq.(8.31) in p.156  of
\citet{VARBOOK}):
\begin{equation}\label{eq7}
\kappa^d=\frac{1}{B} - \frac{ \left( \frac{\partial^2 B}{\partial
P^2} \right)_T } {\left( \frac{\partial B}{\partial P}
\right)_T-1}.
\end{equation}
Since $\left( \frac{\partial^2 B}{\partial P^2} \right)_T$ is
negative (see p.157 of \citet{VARBOOK}), Eq.(\ref{eq7})
indicates that $\kappa^d > \frac{1}{B}$, i.e., $B>B^d$.

\section{Application of the alternative model to Ir-Os alloys}
Equation (\ref{eq4}) can be alternatively written as
\begin{equation}\label{eq8}
\left( \frac{V_{N+n}}{N+n} \right) \frac{1}{B} = \frac{V_1}{N}
\frac{1}{B_1}+  \frac{v^d}{B^d} \left( \frac{n}{N+n} \right)
\end{equation}
where $\frac{V_{N+n}}{N+n}$ for the alloys denotes the mean volume
per atom given in Table 3 of \citet{Yusenko2015155}. This equation
reveals that when plotting $\left( \frac{V_{N+n}}{N+n} \right)
\frac{1}{B}$ versus $\frac{n}{N+n}(=x$, see Eq.(\ref{eq3})) and
making a least square fitting to a straight line the slope leads
to $\frac{v^d}{B^d}$ -from which $B^d$ is calculated since $v^d$
may be approximated by means of Eq.(\ref{eq6})- and the intercept
results in $\frac{V_1}{N} \frac{1}{B_1}$ from which the bulk
modulus $B_1$   of the pure component 1 is determined.

We follow the description of the previous Section and, for the
sake of convenience, we assume as pure components 1 and 2 the
elements Os and Ir (since the volume per atom in Os is smaller
than that in Ir), thus we consider the alloys Ir$_x$Os$_{1-x}$.
\citet{Yusenko2015155} studied four compositions, i.e., $x=0.20,
0.40, 0.55$ and 0.80, and in their Table 3 reported the following
bulk moduli data $B=420(5), 403(16), 393 (7)$, and $368(4)$GPa,
respectively. The mean volumes per atom for the two pure
components are \citep{Yusenko2015155} $v_1=13.9825(1)\times 10^{-30}
m^3$ and $v_2=14.1556(1)\times 10^{-30} m^3$ and hence
$v^d=0.1731(1)\times 10^{-30} m^3$. Using these data, we now plot
in Fig.1 with open red circles the values of $\left(
\frac{V_{N+n}}{N+n} \right) \frac{1}{B}$ versus
$\frac{n}{N+n}(=x)$ and make a weighted least squares
fit (see pp. 659-660 of \citet{numrec}).

\begin{figure}
\noindent\includegraphics[width=20pc]{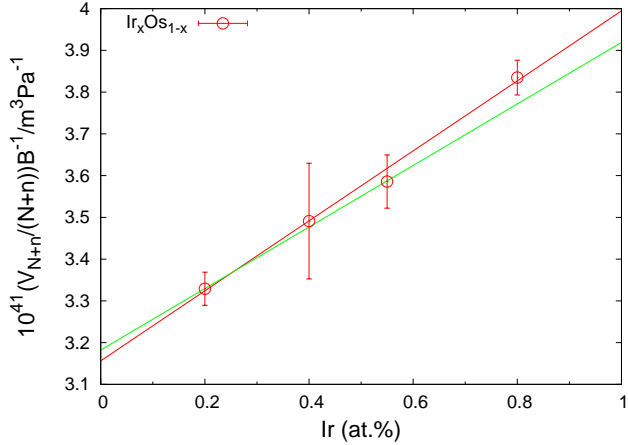}
\caption{The composition dependence of the quantity $\left(
\frac{V_{N+n}}{N+n} \right) \frac{1}{B}$ for the Ir$_x$Os$_{1-x}$
alloys studied by \citet{Yusenko2015155}. The red straight line
corresponds to all four alloys (red circles) according to
Eq.(\ref{eq8}), while in the green straight line the highest Ir
concentration $x=0.80$ is not considered (see the text).}
\end{figure}

The following results have been obtained: When employing all four
alloys studied by \citet{Yusenko2015155} we deduce the red
straight line, the slope of which leads to $B^d=20.6(2.3)$GPa
whereas its intercept to the bulk modulus of Os $B_1=443(8)$GPa
and the corresponding ordinate for $x=1$ to the bulk modulus of
Ir, $B_2=354(10)$GPa. If we do not consider the alloy with the
highest Ir concentration, i.e., the one with $x=0.80$ (for which
the concept of the ``defects'' introduced by Ir when considering
Os matrix may not be fully justified) and restrict ourselves to
the three alloys having the lower $x$ values, i.e., $x=0.20,
0.40,$ and 0.55, we deduce the green straight line from which
slightly different values are obtained. In particular, the slope
leads to $B^d=23(7)$GPa, its intercept to $B_1=439(10)$GPa and the
value $x=1$ corresponds to $B_2=361(21)$GPa. Thus, our results of
the application of the alternative model to Ir$_x$Os$_{1-x}$
alloys could be summarized as follows: The bulk modulus $B_1$ for
Os should lie between 439(10)GPa and 443(8)GPa, while that ($B_2$)
of Ir between 354(10)GPa and 361(21)GPa (cf. as for the $B^d$
values they were found appreciably smaller than $B$ as expected
from the thermodynamical model discussed in the previous Section).
In other words, our findings concerning the bulk modulus of Os do
not support the claim by \citet{PhysRevLett.88.135701} that it is
markedly larger than the bulk modulus reported by \cite{BOs} for
diamond which is 443GPa. As for our findings concerning the bulk
modulus $B_2$ of Ir, we see that it lies between the values
354(6)GPa and 391(4)GPa reported by
\citet{Cerenius200026} and
\citet{PhysRevB.75.224114}, respectively.

\section{Summary and Conclusions}
Here, we made use of the bulk moduli of fcc- and hcp-structured
Ir-Os alloys recently measured by \citet{Yusenko2015155} in order
to estimate the bulk modulus of osmium for which controversial
values have been reported. We find that it lies between 439(10)GPa
and 443(8)GPa by employing a model different than that used in
\citep{Yusenko2015155}. Our results strengthen the
conclusion
that the
bulk modulus for Os is slightly smaller in comparison with diamond
($B=$443GPa). Furthermore, our findings contradict the earlier
claim of \citet{PhysRevLett.88.135701} that osmium is less
compressible than diamond, which is particularly interesting due
to the significance of Os as a possible model of iron behavior in
the Earth's core.


%
%
%
%
%
%
%

\section*{Acknowledgments}
The experimental data used in the present paper come from Tables 2 and 3 of \citet{Yusenko2015155}.



\begin{thebibliography}{49}
\providecommand{\natexlab}[1]{#1}
\expandafter\ifx\csname urlstyle\endcsname\relax
  \providecommand{\doi}[1]{doi:\discretionary{}{}{}#1}\else
  \providecommand{\doi}{doi:\discretionary{}{}{}\begingroup
  \urlstyle{rm}\Url}\fi

\bibitem[{\textit{Alexopoulos and Varotsos}(1981)}]{PhysRevB.24.3606}
Alexopoulos, K., and P.~Varotsos (1981), {Calculation of diffusion coefficients
  at any temperature and pressure from a single measurement. II.
  Heterodiffusion}, \textit{Phys. Rev. B}, \textit{24}, 3606--3609,
  \doi{10.1103/PhysRevB.24.3606}.

\bibitem[{\textit{Armentrout and Kavner}(2010)}]{ARM10}
Armentrout, M.~M., and A.~Kavner (2010), Incompressibility of osmium metal at
  ultrahigh pressures and temperatures, \textit{J. Appl. Phys.},
  \textit{107}(9), 093,528, \doi{10.1063/1.3369283}.

\bibitem[{\textit{Bird and Bassett}(1980)}]{BIR80}
Bird, J.~M., and W.~A. Bassett (1980), Evidence of a deep mantle history in
  terrestrial osmium-iridium-ruthenium alloys, \textit{J. Geophys. Res.: Solid
  Earth}, \textit{85}, 2156--2202, \doi{10.1029/JB085iB10p05461}.

\bibitem[{\textit{Cerenius and Dubrovinsky}(2000)}]{Cerenius200026}
Cerenius, Y., and L.~Dubrovinsky (2000), Compressibility measurements on
  iridium, \textit{Journal of Alloys and Compounds}, \textit{306}(1-2), 26 --
  29, \doi{10.1016/S0925-8388(00)00767-2}.

\bibitem[{\textit{Chroneos and Vovk}(2015{\natexlab{a}})}]{Chroneos20151}
Chroneos, A., and R.~Vovk (2015{\natexlab{a}}), {Modeling self-diffusion in
  UO$_2$ and ThO$_2$ by connecting point defect parameters with bulk
  properties}, \textit{Solid State Ionics}, \textit{274}, 1 -- 3,
  \doi{http://dx.doi.org/10.1016/j.ssi.2015.02.010}.

\bibitem[{\textit{Chroneos and Vovk}(2015{\natexlab{b}})}]{Chroneos2015179}
Chroneos, A., and R.~Vovk (2015{\natexlab{b}}), {Connecting bulk properties of
  germanium with the behavior of self- and dopant diffusion}, \textit{Materials
  Science in Semiconductor Processing}, \textit{36}, 179 -- 183,
  \doi{http://dx.doi.org/10.1016/j.mssp.2015.03.053}.

\bibitem[{\textit{Creager}(1992)}]{CRE92}
Creager, K.~C. (1992), Anisotropy of the inner core from differential travel
  times of the phases, \textit{Nature (London)}, \textit{356}, 309,
  \doi{10.1038/356309a0}.

\bibitem[{\textit{Cynn et~al.}(2002)\textit{Cynn, Klepeis, Yoo, and
  Young}}]{PhysRevLett.88.135701}
Cynn, H., J.~E. Klepeis, C.-S. Yoo, and D.~A. Young (2002), Osmium has the
  lowest experimentally determined compressibility, \textit{Phys. Rev. Lett.},
  \textit{88}, 135,701, \doi{10.1103/PhysRevLett.88.135701}.

\bibitem[{\textit{Fonseca et~al.}(2012)\textit{Fonseca, Laurenz, Mallmann,
  Luguet, Hoehne, and Jochum}}]{FON12}
Fonseca, R.~O., V.~Laurenz, G.~Mallmann, A.~Luguet, N.~Hoehne, and K.~P. Jochum
  (2012), {New constraints on the genesis and long-term stability of Os-rich
  alloys in the Earth's mantle}, \textit{Geochim. Cosmochim. Acta},
  \textit{87}, 227 -- 242, \doi{10.1016/j.gca.2012.04.002}.

\bibitem[{\textit{Geballe et~al.}(1963)\textit{Geballe, Matthias, Compton,
  Corenzwit, and Hull}}]{GEB63}
Geballe, T.~H., B.~T. Matthias, V.~B. Compton, E.~Corenzwit, and G.~W. Hull
  (1963), Superconductivity of solid solutions of noble metals, \textit{Phys.
  Rev.}, \textit{129}, 182--183, \doi{10.1103/PhysRev.129.182}.

\bibitem[{\textit{Godwal et~al.}(2012)\textit{Godwal, Yan, Clark, and
  Jeanloz}}]{GOD12}
Godwal, B.~K., J.~Yan, S.~M. Clark, and R.~Jeanloz (2012), High-pressure
  behavior of osmium: An analog for iron in earth's core, \textit{J. Appl.
  Phys.}, \textit{111}(11), 112,608, \doi{10.1063/1.4726203}.

\bibitem[{\textit{Goncharov et~al.}(2007)\textit{Goncharov, Crowhurst,
  Dewhurst, Sharma, Sanloup, Gregoryanz, Guignot, and
  Mezouar}}]{PhysRevB.75.224114}
Goncharov, A.~F., J.~C. Crowhurst, J.~K. Dewhurst, S.~Sharma, C.~Sanloup,
  E.~Gregoryanz, N.~Guignot, and M.~Mezouar (2007), {Thermal equation of state
  of cubic boron nitride: Implications for a high-temperature pressure scale},
  \textit{Phys. Rev. B}, \textit{75}, 224,114,
  \doi{10.1103/PhysRevB.75.224114}.

\bibitem[{\textit{Kenichi}(2004)}]{TAK04}
Kenichi, T. (2004), Bulk modulus of osmium: High-pressure powder x-ray
  diffraction experiments under quasihydrostatic conditions, \textit{Phys. Rev.
  B}, \textit{70}, 012,101, \doi{10.1103/PhysRevB.70.012101}.

\bibitem[{\textit{Morelli et~al.}(1986)\textit{Morelli, Dziewonski, and
  Woodhouse}}]{MORE86}
Morelli, A., A.~M. Dziewonski, and J.~H. Woodhouse (1986), Anisotropy of the
  inner core inferred from pkikp travel times, \textit{Geophys. Res. Lett.},
  \textit{13}, 1545, \doi{10.1029/GL013i013p01545}.

\bibitem[{\textit{Occelli et~al.}(2003)\textit{Occelli, Loubeyre, and
  LeToullec}}]{BOs}
Occelli, F., P.~Loubeyre, and R.~LeToullec (2003), {Properties of diamond under
  hydrostatic pressures up to 140 GPa}, \textit{Nat. Mater.}, \textit{2},
  151--154, \doi{10.1038/nmat831}.

\bibitem[{\textit{Okamoto}(1994)}]{OKA94}
Okamoto, H. (1994), {The Ir-Os (iridium-osmium) system}, \textit{J. Phase
  Equilib.}, \textit{15}, 55--57, \doi{10.1007/BF02667683}.

\bibitem[{\textit{Press et~al.}(1992)\textit{Press, Teukolsky, Vetterling, and
  Flannery}}]{numrec}
Press, W.~H., S.~Teukolsky, W.~Vetterling, and B.~P. Flannery (1992),
  \textit{Numerical Recipes in FORTRAN}, {963} pp., {Cambridge University
  Press}, {New York}.

\bibitem[{\textit{Reiswig and Dickinson}(1964)}]{REI64}
Reiswig, R.~D., and J.~M. Dickinson (1964), The osmium-iridium equilibrium
  diagram, \textit{Trans. Metall. Soc. AIME}, \textit{230}, 469--472.

\bibitem[{\textit{Sarlis et~al.}(2010)\textit{Sarlis, Skordas, and
  Varotsos}}]{NEWTSA}
Sarlis, N.~V., E.~S. Skordas, and P.~A. Varotsos (2010), Nonextensivity and
  natural time: The case of seismicity, \textit{Phys. Rev. E}, \textit{82},
  021,110, \doi{10.1103/PhysRevE.82.021110}.

\bibitem[{\textit{Skordas}(2012)}]{SKO12}
Skordas, E.~S. (2012), Comments on the elastic properties in solid solutions of
  silver halides, \textit{Mod. Phys. Lett. B}, \textit{26}, 1250,066,
  \doi{10.1142/S0217984912500662}.

\bibitem[{\textit{Skordas et~al.}(2010)\textit{Skordas, Sarlis, and
  Varotsos}}]{CHAOS2010}
Skordas, E.~S., N.~V. Sarlis, and P.~A. Varotsos (2010), Effect of significant
  data loss on identifying electric signals that precede rupture estimated by
  detrended fluctuation analysis in natural time, \textit{CHAOS}, \textit{20},
  033,111, \doi{10.1063/1.3479402}.

\bibitem[{\textit{Vacher et~al.}(1954)\textit{Vacher, Bechtoldt, and
  Maxwell}}]{VAC54}
Vacher, H.~C., C.~J. Bechtoldt, and E.~Maxwell (1954), Structure of some
  osmium-iridium alloys, \textit{J. Metal.}, \textit{200}, 80--82.

\bibitem[{\textit{Varotsos}(1974)}]{PhysRevB.9.1866}
Varotsos, P. (1974), {Conductivity and dielectric constants of LiD},
  \textit{Phys. Rev. B}, \textit{9}, 1866--1869, \doi{10.1103/PhysRevB.9.1866}.

\bibitem[{\textit{Varotsos}(1978)}]{VARO78}
Varotsos, P. (1978), An estimate of the pressure dependence of the dielectric
  constant in alkali halides, \textit{Phys. Stat. Sol. (b)}, \textit{90},
  339--343.

\bibitem[{\textit{Varotsos}(1980{\natexlab{a}})}]{VAR80AA}
Varotsos, P. (1980{\natexlab{a}}), On the temperature variation of the bulk
  modulus of mixed alkali halides, \textit{Phys. Status Sol. B},
  \textit{99}(2), K93--K96, \doi{10.1002/pssb.2220990243}.

\bibitem[{\textit{Varotsos}(1980{\natexlab{b}})}]{PSSB:PSSB2221000252}
Varotsos, P. (1980{\natexlab{b}}), Determination of the dielectric constant of
  alkali halide mixed crystals, \textit{physica status solidi (b)},
  \textit{100}(2), K133--K138, \doi{10.1002/pssb.2221000252}.

\bibitem[{\textit{Varotsos}(1981)}]{VAR81}
Varotsos, P. (1981), Determination of the composition of the maximum
  conductivity or diffusivity in mixed alkali halides, \textit{J. Phys. Chem.
  Sol.}, \textit{42}, 405--407, \doi{10.1016/0022-3697(81)90048-2}.

\bibitem[{\textit{Varotsos}(2008)}]{Varotsos2008438}
Varotsos, P. (2008), {Point defect parameters in $\beta$-PbF$_{2}$ revisited},
  \textit{Solid State Ionics}, \textit{179}(11?12), 438 -- 441,
  \doi{10.1016/j.ssi.2008.02.055}.

\bibitem[{\textit{Varotsos and Alexopoulos}(1977)}]{VAR77A}
Varotsos, P., and K.~Alexopoulos (1977), Calculation of the formation entropy
  of vacancies due to anharmonic effects, \textit{Phys. Rev. B}, \textit{15},
  4111--4114, \doi{10.1103/PhysRevB.15.4111}.

\bibitem[{\textit{Varotsos and
  Alexopoulos}(1980{\natexlab{a}})}]{PSSB:PSSB2221020157}
Varotsos, P., and K.~Alexopoulos (1980{\natexlab{a}}), {Determination of the
  Compressibility of an Alloy from Its Density}, \textit{physica status solidi
  (b)}, \textit{102}(1), K67--K72, \doi{10.1002/pssb.2221020157}.

\bibitem[{\textit{Varotsos and
  Alexopoulos}(1980{\natexlab{b}})}]{doi:10.1080/01418618008239352}
Varotsos, P., and K.~Alexopoulos (1980{\natexlab{b}}), On the question of the
  calculation of migration volumes in ionic crystals, \textit{Philosophical
  Magazine A}, \textit{42}(1), 13--18, \doi{10.1080/01418618008239352}.

\bibitem[{\textit{Varotsos and Alexopoulos}(1980{\natexlab{c}})}]{VARALEX80}
Varotsos, P., and K.~Alexopoulos (1980{\natexlab{c}}), Migration entropy for
  the bound fluorine motion in alkaline earth fluorides, \textit{J. Phys. Chem.
  Sol.}, \textit{41}, 443--446, \doi{10.1016/0022-3697(80)90172-9}.

\bibitem[{\textit{Varotsos and Alexopoulos}(1980{\natexlab{d}})}]{VAR80B}
Varotsos, P., and K.~Alexopoulos (1980{\natexlab{d}}), Prediction of the
  compressibility of mixed alkali halides, \textit{Journal of Physics and
  Chemistry of Solids}, \textit{41}(12), 1291 -- 1294.

\bibitem[{\textit{Varotsos and Alexopoulos}(1981)}]{VAR819}
Varotsos, P., and K.~Alexopoulos (1981), {Migration parameters for the bound
  fluorine motion in alkaline earth fluorides. II}, \textit{J. Phys. Chem.
  Sol.}, \textit{42}, 409 -- 410, \doi{10.1016/0022-3697(81)90049-4}.

\bibitem[{\textit{Varotsos and Alexopoulos}(1986)}]{VARBOOK}
Varotsos, P., and K.~Alexopoulos (1986), \textit{Thermodynamics of Point
  Defects and their Relation with Bulk Properties}, 474 pp., North Holland,
  Amsterdam.

\bibitem[{\textit{Varotsos and Miliotis}(1974)}]{VAR74B}
Varotsos, P., and S.~Miliotis (1974), New aspects on the dielectric properties
  of the alkali halides with divalent impurities, \textit{J. Phys. Chem. Sol.},
  \textit{35}, 927--930.

\bibitem[{\textit{Varotsos and Mourikis}(1974)}]{VAR74}
Varotsos, P., and S.~Mourikis (1974), Difference in conductivity between lid
  and lih crystals, \textit{Phys. Rev. B}, \textit{10}, 5220--5224.

\bibitem[{\textit{Varotsos et~al.}(1978)\textit{Varotsos, Ludwig, and
  Alexopoulos}}]{VAR78B}
Varotsos, P., W.~Ludwig, and K.~Alexopoulos (1978), Calculation of the
  formation volume of vacancies in solids, \textit{Phys. Rev. B}, \textit{18},
  2683--2691, \doi{10.1103/PhysRevB.18.2683}.

\bibitem[{\textit{Varotsos et~al.}(1985)\textit{Varotsos, Alexopoulos,
  Varotsos, and Lazaridou}}]{PSSA:PSSA2210880250}
Varotsos, P., K.~Alexopoulos, C.~Varotsos, and M.~Lazaridou (1985),
  {Interconnection of point defect parameters in BaF$_2$}, \textit{physica
  status solidi (a)}, \textit{88}(2), K137--K140,
  \doi{10.1002/pssa.2210880250}.

\bibitem[{\textit{Varotsos et~al.}(1992)\textit{Varotsos, Bogris, and
  Kyritsis}}]{VAR92}
Varotsos, P., N.~Bogris, and A.~Kyritsis (1992), Comments on the depolarization
  currents stimulated by variations of temperature and pressure, \textit{J.
  Phys. Chem. Solids}, \textit{53}, 1007--1011,
  \doi{10.1016/0022-3697(92)90069-P}.

\bibitem[{\textit{Varotsos}(2006)}]{VAR06B}
Varotsos, P.~A. (2006), What happened before the last five strong earthquakes
  in greece, \textit{Proc. Jpn. Acad., Ser. B: Phys. Biol. Sci.}, \textit{82},
  86--91, \doi{10.2183/pjab.82.86}.

\bibitem[{\textit{Varotsos}(2007)}]{PhysRevB.75.172107}
Varotsos, P.~A. (2007), Calculation of point defect parameters in diamond,
  \textit{Phys. Rev. B}, \textit{75}, 172,107,
  \doi{10.1103/PhysRevB.75.172107}.

\bibitem[{\textit{Varotsos et~al.}(2008)\textit{Varotsos, Sarlis, Skordas, and
  Lazaridou}}]{NAT08}
Varotsos, P.~A., N.~V. Sarlis, E.~S. Skordas, and M.~S. Lazaridou (2008),
  Fluctuations, under time reversal, of the natural time and the entropy
  distinguish similar looking electric signals of different dynamics,
  \textit{J. Appl. Phys.}, \textit{103}, 014906, \doi{10.1063/1.2827363}.

\bibitem[{\textit{Varotsos et~al.}(2009)\textit{Varotsos, Sarlis, and
  Skordas}}]{NAT09V}
Varotsos, P.~A., N.~V. Sarlis, and E.~S. Skordas (2009), Detrended fluctuation
  analysis of the magnetic and electric field variations that precede rupture,
  \textit{CHAOS}, \textit{19}, 023,114, \doi{10.1063/1.3130931}.

\bibitem[{\textit{Varotsos et~al.}(1999)\textit{Varotsos, Sarlis, and
  Lazaridou}}]{VAR99}
Varotsos, P.~V., N.~V. Sarlis, and M.~S. Lazaridou (1999), Interconnection of
  defect parameters and stress-induced electric signals in ionic crystals,
  \textit{Phys. Rev. B}, \textit{59}, 24--27, \doi{10.1103/PhysRevB.59.24}.

\bibitem[{\textit{Voronin et~al.}(2005)\textit{Voronin, Pantea, Zerda, Wang,
  and Zhao}}]{VOR05}
Voronin, G., C.~Pantea, T.~Zerda, L.~Wang, and Y.~Zhao (2005), Thermal
  equation-of-state of osmium: a synchrotron x-ray diffraction study,
  \textit{Journal of Physics and Chemistry of Solids}, \textit{66}(5), 706 --
  710, \doi{10.1016/j.jpcs.2004.08.045}.

\bibitem[{\textit{Weinberger et~al.}(2008)\textit{Weinberger, Tolbert, and
  Kavner}}]{WEI08}
Weinberger, M.~B., S.~H. Tolbert, and A.~Kavner (2008), Osmium metal studied
  under high pressure and nonhydrostatic stress, \textit{Phys. Rev. Lett.},
  \textit{100}, 045,506, \doi{10.1103/PhysRevLett.100.045506}.

\bibitem[{\textit{Yusenko et~al.}(2015)\textit{Yusenko, Bykova, Bykov,
  Gromilov, Kurnosov, Prescher, Prakapenka, Hanfland, van Smaalen, Margadonna,
  and Dubrovinsky}}]{Yusenko2015155}
Yusenko, K.~V., E.~Bykova, M.~Bykov, S.~A. Gromilov, A.~V. Kurnosov,
  C.~Prescher, V.~B. Prakapenka, M.~Hanfland, S.~van Smaalen, S.~Margadonna,
  and L.~S. Dubrovinsky (2015), {Compressibility of Ir-Os alloys under high
  pressure}, \textit{J. Alloys Compd.}, \textit{622}, 155--161,
  \doi{10.1016/j.jallcom.2014.09.210}.

\bibitem[{\textit{Zhang and Shan}(2015)}]{GGGE:GGGE20681}
Zhang, B., and S.~Shan (2015), {Application of the cB$\Omega$ model to the
  calculation of diffusion parameters of Si in silicates},
  \textit{Geochemistry, Geophysics, Geosystems}, \textit{16}(3), 705--718,
  \doi{10.1002/2014GC005551}.

\end{thebibliography}
\end{document}